# Ultra-high Hydrogen Storage Capacity of Holey Graphyne

Yan Gao[a,b], Huanian Zhang[c], Hongzhe Pan[d], Qingfang Li[e], Haifeng Wang[a] and Jijun Zhao[b,*]

[a]Department of Physics, College of Science, Shihezi University, Xinjiang 832003, China

[b]Key Laboratory of Materials Modification By Laser, Ion and Electron Beams (Dalian University of Technology), Ministry of Education, Dalian 116024, China

[c]School of Physics & Optoelectronic Engineering, Shandong University of Technology, Zibo 255049, China

[d]School of Physics and Electronic Engineering, Linyi University, Linyi 276005, China

[e]School of Physics & Optoelectronic Engineering, Nanjing University of Information Science & Technology, Nanjing 210044, China

**ABSTRACT:** Holey graphyne (HGY), a novel 2D single-crystalline carbon allotrope, was synthesized most recently by Castro-Stephens coupling reaction. The natural existing uniform periodic holes in the 2D carbon-carbon network demonstrate its tremendous potential application in the area of energy storage. Herein, we conducted density functional theory (DFT) calculation to predict the hydrogen storage capacity of HGY sheet. It's found the Li-decorated single-layer HGY can serve as a promising candidate for hydrogen storage. Our numerical calculations demonstrate that Li atoms can bind strongly to the HGY sheet without the formation of Li clusters, and each Li atom can anchor four $H_2$ molecules with the average adsorption energy about -0.22 eV/$H_2$. The largest hydrogen storage capacity of the doped HGY sheet can arrive as high as 12.8 wt%, this value largely surpasses the target of the U. S. Department of Energy (9 wt%), showing the Li/HGY complex is an ideal hydrogen storage material at ambient conditions. In addition, we investigate the polarization mechanism of the storage media and and find that the polarization stemed from both the electric field induced by the ionic Li decorated on the HGY and the weak polarized hydrogen molecules dominated the $H_2$ adsorption process.

--------------------------

Corresponding author. E-mail: zhaojj@dlut.edu.cn (J. Z.).

# 1. Introduction

Nowadays, the exhaustion of fossil fuel has stimulated numerous studies on the search of renewable and sustainable energy resources. As an ideal alternative clean energy source, hydrogen energy has many advantages, such as lightweight, highly abundant and zero production of greenhouse gaseous from the combustion [1]. The biggest challenges to make full use of the hydrogen energy is the high density hydrogen storage. It's specified by the U. S. Department of Energy (DOE) in 2015 that, to perform reversibly hydrogen charging cycles at ambient conditions for automobile on board applications, a desired storage system should have sufficiently high gravimetric density (>9 wt %), otherwise, the binding energy of hydrogen molecule should lie in a suitable energy range of -0.2 to -0.4 eV per $H_2$ [2,3].

Benefiting of large surface area and lightweight, two-dimentional (2D) materials are suggested to be the promising hydrogen storage medium in recent years. Till now, a great deal of 2D materials, e.g., graphene [4], h-BN [5], $MoS_2$ [6], black phosphorene [7] and holey-$C_2N$ [8] are explored to be promising materials for hydrogen storage. Among of various 2D materials, graphyne (GY) [9], a theorized one-atom thick carbon allotrope consist of $sp^2$ hybridized carbons (benzene rings) bridged by ethylenic groups ($sp$ carbons), is expected to possess many unique properties, which brings about tremendous application potentials in the fields of electronics, optoelectronics, catalysis, sensor and energy [10-16]. In experiment, although many analogous structures of GY, e.g., GDY [17,18], BGDY [19], H-, F- and Me-GY [20] have been synthesized during the past years, the true GY, in which only one acetylenic bond between adjacent benzene rings, has not been created successfully due to the synthetic challenge. Until very recently, a novel true GY material, 2D single-crystalline holey-graphyne (HGY) was fabricated by Castro-Stephens coupling reaction from 1,3,5-tribromo-2,4,6-triethynyl benzene [21]. The carbon-carbon 2D network is alternately connected between benzene rings and $sp$ bonds.

The most recent success in synthesis of HGY nanosheet provides an updated account of the scientific and applied interest and prospect of catalysis, sensor and energy fields. Importantly, due to the natural existing uniform holes in the geometrical structure of HGY sheet, it may be a good candidate for energy storage. In this paper, we conduct an ab initio prediction of its potential in hydrogen storage. It's found the pristine HGY sheet without any extra atoms is not suitable to capture hydrogen molecules because of weak adsorption energy. While Li-decorated HGY can result in a medium of ultra-high capacity hydrogen storage of 12.8 wt % with an average adsorption energy of about -0.22 eV/$H_2$. Our results suggest a potential host material for hydrogen storage.

## 2. Computational details

All the first-principles calculations were performed by using the density functional theory as implemented in the Vienna ab-initio simulation package (VASP) [22,23] with the projector-augmentedwave (PAW) method. The Perdew-Burke-Ernzerhof (PBE) [24] of generalized gradient approximation up to 500 eV was chosen as the exchange-correlation functional. A Monkhorst-Pack k-mesh of 5×5×1 was used to sample the Brillouin zone in the structure optimization for single-layer HGY. The convergence for energy was chosen as $10^{-5}$ eV between two steps. Both the lattice constants and internal coordinates were optimized until the Hellman-Feynman forces on each atom was less than $10^{-2}$ eV Å$^{-1}$. A vacuum space of 20 Å was taken to safely avoid the interactions in the non-periodic directions. The optimized lattice constants for the HGY monolayer are a=b=10.85 Å, and a supercell composed 2×2 unit cells has been chosen for the calculations of hydrogen storage on HGY. In our treatment of van der Waals correction, the empirical correction scheme of Grimme (DFT+D2) [25] has been employed. In order to check the stability of storage materials, we performed the ab initio molecular dynamics at 300 K up to 5 ps with the time-step 0.5 fs.

We define the binding energy ($E_b$) of the HGY monolayer with Li atoms using the expression,

$$E_b = \frac{1}{n}(E_{HGY+nLi} - E_{HGY} - nE_{Li}), \quad (1)$$

where $E_{HGY+nLi}$ and $E_{HGY}$ are the total energies of HGY monolayer with and without Li adsorption, respectively. $E_{Li}$ and $n$ are the energy of an isolated Li atom and the number of Li atoms, respectively.

The average adsorption energy $E_{ads}$ of H$_2$ molecules absorbed on Li-decorated HGY are investigated using the relation:

$$E_{ads} = \frac{1}{m}(E_{HGY+nLi+mH_2} - E_{HGY+nLi} - mE_{H_2}), \qquad (2)$$

where $E_{HGY+nLi+mH_2}$ is the total energy of Li-decorated HGY sheet with $m$ H$_2$ molecules. $E_{H_2}$ represents the energy of an isolated hydrogen molecule, $m$ is the numbers of H$_2$ molecules.

## 3. Results and discussion

The geometrical structure of HGY sheet is shown in Fig. 1. For comparison, we also plotted the structure of GY [9] in Fig. S1 of the Supporting Information (SI). Single-layer HGY adopts a graphene-like hexagonal lattice with space group P6/mmm. In each unit cell of HGY, there are total 24 carbon atoms and can be divided into two different types (C$_1$ and C$_2$, see Fig. 1). While there are four types of C-C bonds: including two *sp$^2$-sp$^2$* bonds (B$_1$=1.463Å, B$_2$=1.397Å), one *sp-sp$^2$* bond (B$_3$=1.414Å) and one *sp-sp* bond (B$_4$=1.227Å). HGY sheet can be considered as the hexagonal benzene rings joined together by acetylenic linkages ($-C \equiv C-$), which is similar with GY (Fig. S1). The major difference between GY and HGY is that, in structures of GY, three acetylenic linkages surround to form a large triangle pore, and the angles between acetylenic linkages and benzene rings are exactly 120°, while in HGY, six acetylenic linkages surround to form a large pore and the angles between acetylenic linkages and benzene rings are about 125.8°. Besides, two adjacent acetylenic linkages in HGY concurrently join to the neighboring benzene rings resulting in an eight-vertex ring. Our optimized lattice constants of HGY are a=b=10.85 Å, agrees well with the experiment report [21].

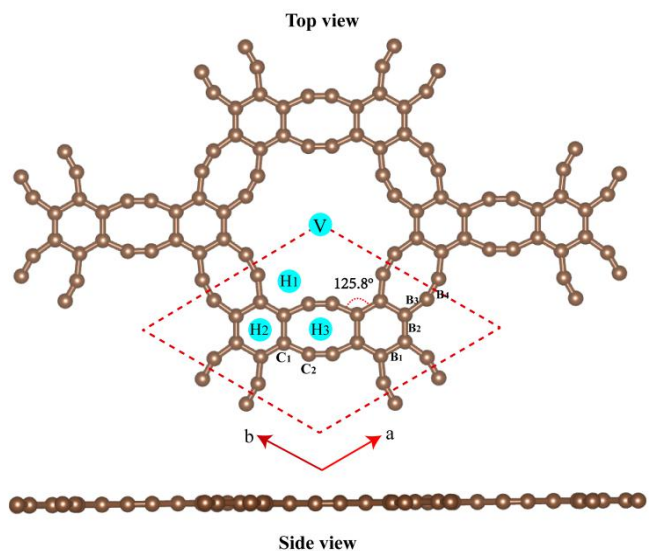

**Fig. 1.** Top and side views of the atomic configuration of single-layer HGY sheet. The unit cell is indicated by the red dashed lines. The $C_1$, $C_2$ denote the different carbon atoms, $B_1$, $B_2$, $B_3$, $B_4$ denote the different types of C-C bonds, and V, $H_1$, $H_2$, $H_3$ represent different hole sites in HGY.

The uniformly distributed holes in HGY sheet, including middle large pores, six-carbon benzene rings and eight-vertex rings may bring about great potential in energy storage, herein, we give a prediction of the hydrogen storage capacity of HGY sheet based on DFT calculations. Firstly, the hydrogen storage property of the pristine HGY sheet should be evaluated. To find the most stable adsorption site of single hydrogen molecule, we considered ten different high symmetry sites, including four hole sites (V, $H_1$, $H_2$ and $H_3$), two top sites ($C_1$-, $C_2$-top) and four bridge sites ($B_1$, $B_2$, $B_3$ and $B_4$). The calculated absorption energies for the above possible sites are presented in Table S1 of SI, and the most energy favorable configuration is illustrated in Fig. S2. It's found that the largest absorption energies for a single hydrogen molecule on the pristine HGY is only -0.06 eV, suggesting that $H_2$ adsorption on the pristine HGY sheet without any extra element was too weak to ensure an efficient hydrogen storage.

In order to enhance the the adsorption ability of HGY sheet, Li incorporated system is considered. We begin with checking the adsorption behavior of a single Li atom decorated on the HGY sheet. All the possible adsorption sites on available symmetry positions are investigated, and the corresponding calculated binding energies are presented in Table I. It's found that only the hole sites can capture Li atom, when a Li atom was initially placed at the C top sites or C-C bridge sites, it would move to the $H_3$-hole sites. Besides, the middle large hole (V-site) is also found unable to capture single Li atom, it would move to the corner of two neighbouring acetylenic linkages and benzene rings, i.e., $H_1$ hole site, with the binding energy of -2.42 eV. The binding energy of $H_2$-site (the top of benzene rings) is calculated as -2.22 eV. The most stable site is $H_3$ hole site, i.e., the top of eight-vertex carbon rings with the binding energy -2.47 eV, the obtained configuration is illustrated in Fig. 2a. All the binding energies of the three situations are significantly lower than the cohesive energy of bulk Li (-1.795 eV/atom).26 Thus, the aggregation of Li atoms in the HGY monolayer can be neglected, confirming the stability of Li-decorated HGY.

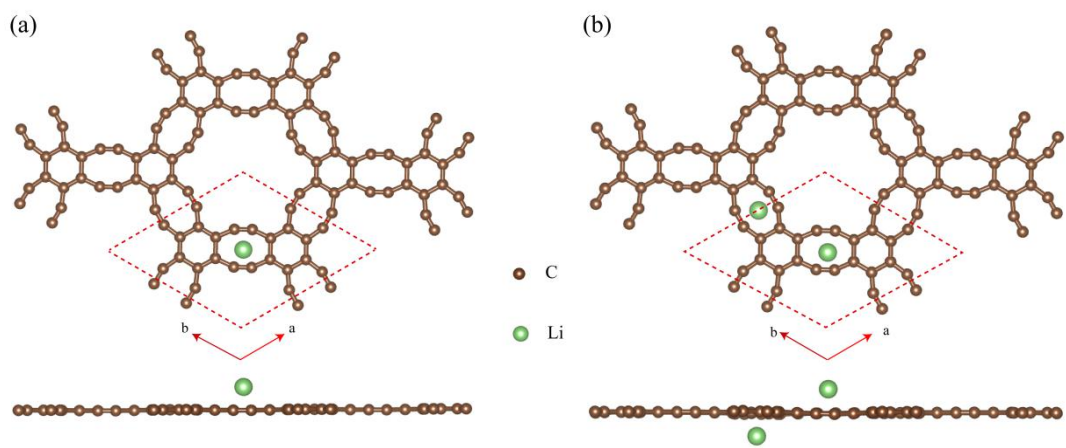

**Fig. 2.** Configurations of the most energy favorable (a) single Li atom and (b) two Li atoms decorated on HGY.

**Table 1.** Calculated Binding Energies (in eV) of Single Li Atom Decorated in HGY Sheet.

| System | Top | | Bridge | | | | Hole | | | |
|---|---|---|---|---|---|---|---|---|---|---|
| Sites | $C_1$ | $C_2$ | $B_1$ | $B_2$ | $B_3$ | $B_4$ | V | $H_1$ | $H_2$ | $H_3$ |
| Binding energy | moved to $H_3$-site | | moved to $H_3$-site | | | | moved to $H_1$-site | -2.41 | -2.22 | -2.47 |

We then continue to increase Li adatoms onto the HGY sheet. Six most energy favorable configurations for two Li atoms absorbed in HGY monolayer are carefully checked, including two Li atoms adsorbed on the opposite sides of the HGY layer at the same $H_1$, $H_2$, $H_3$-sites (Fig. S3. a,b and c), two Li atoms placed on two adjacent $H_1$, $H_2$, $H_3$-sites (Fig. S3. d,e and f). The corresponding binding energies are illustrated in Table. SII. The most energy favorable configuration of two Li atoms decorated in HGY is illustrated in Fig. 2b, where two Li atoms were adsorbed in two adjacent $H_3$-sites on both sides of HGY layer. The binding energy of this situation (-2.46 eV/Li) is almost identical to that of single Li atom absorbed on $H_3$-site (-2.47 eV/Li). Therefore, it's clear that with the number of Li atoms increasing, the Li atoms are more prefer to located on the $H_3$-sites.

On the base of the former analysis, we continually add the numbers of decorated Li atoms. It's found that at most six Li atoms can be captured in a unit cell of HGY with three Li atoms located above the sheet and other three below the sheet, as illustrated in Fig. 3a. All the Li atoms are located on the top of the eight-vertex carbon rings with the distance from HGY sheet about 1.52 Å. The average binding energy of one Li atom is calculated as -2.04 eV. Through the bader charge analysis, we found that per Li atom donate about 0.85 |e| to the C substrate. Such charge redistribution induces the coulomb interactions between Li atom and HGY sheet, which is not only beneficial for the stability of 6Li-decorated HGY layer (6Li-HGY), but also may result in a local electric field. In Fig. 3b, we present the charge density difference of 6Li-HGY. It is observed the electrons deplete around Li adatom and accumulate near the C atoms, which leads to the strong electrostatic interaction between the metal atom and HGY. The result is consistent with the above bader charge analysis.

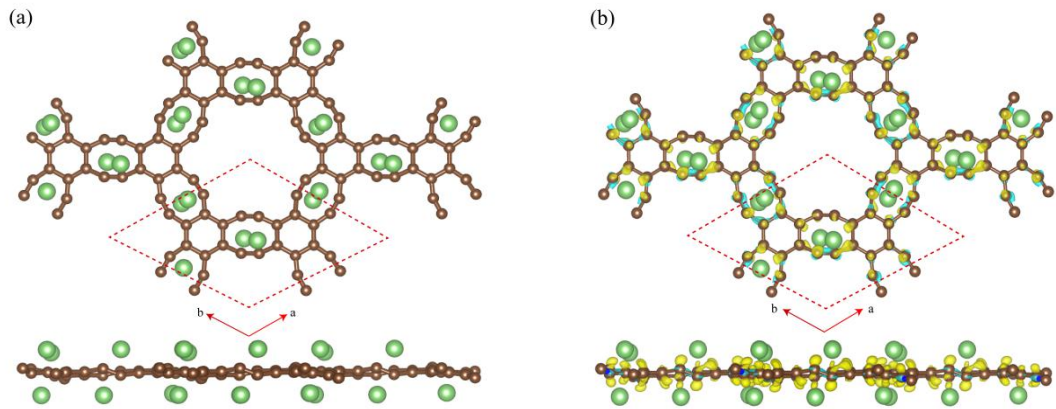

**Fig. 3.** (a) Top and side view of 6Li-HGY sheet. (b) The charge density difference of 6Li-HGY with iso-surface value of $1.0 \times 10^{-2}$ e/Å$^3$. The increase and loss of the electron density are indicated in gold and cyan colors, respectively.

Ab initio molecular dynamics simulations were performed to investigate the thermal stability of 6Li-decorated HGY at 300 K up to 5 ps, the final results are shown in Fig. S4. One can see that the Li atoms were well stabilized on HGY layer and without clustering of Li atoms, only a slight corrugation in the HGY layer was found. The result indicates that this storage media is thermal stable enough and may maintain a good recycle stability.

Now, we turn to discuss the hydrogen adsorption on the 6Li-HGY sheet. We start with a single $H_2$ molecule per one Li atom absorbed on this storage media (6$H_2$ in 6Li-HGY). The most favorable adsorption geometry for this case is shown in Fig. 4a, the hydrogen molecules are found prefer to stick aslant to the Li atoms and tilt slightly to one of the acetylenic linkage in HGY sheet. Then we gradually increased the number of adsorbed $H_2$ molecules on the Li decorated HGY layer. The most favorable adsorption geometry for two $H_2$ molecule per Li case (12$H_2$ in 6Li-HGY) is shown in Fig. 4b. Two $H_2$ molecule symmetrically distributed on both sides of the Li atom and tilt slightly to the two adjacent acetylenic linkages of one eight-vertex rings. Such symmetrically distribution of $H_2$ molecules are more obvious in three $H_2$ molecule per Li case (18$H_2$ in 6Li-HGY), where three $H_2$ molecules formed a nearly regular triangle, see Fig. 4c. This tendency mainly stems from the 3-fold rotational symmetrical potential well induced by the HGY sheet. We found that the maximum of four $H_2$ molecules per Li atom can be adsorbed on the 6Li-HGY layer, as shown in Fig. 4d. However, the last hydrogen molecule has a weak interaction with the Li/HGY complex (about -0.08 eV/$H_2$), more importantly, the distance of the forth $H_2$ molecules from Li atom is much far than that of the former three $H_2$ molecules, as can be seen from Fig. 4d. No more hydrogen molecules can be captured to the storage medium according to our calculations. So, the maximum of 24 $H_2$ molecules can be captured on both sides of the 6Li-HGY layer, which gives an maximum weight percentage of 12.8 wt %, the value can largely surpass the target of the DOE [2,3], and much higher than the values obtained in alkali metal decorated graphene [4], h-BN [5], $MoS_2$ [6] and phosphorene [7] systems.

Table II lists the calculated $H_2$ adsorption details, including the average adsorption, the distances of $H_2$ molecules from Li atoms and H-H bond lengths depending on the numbers of adsorbed $H_2$ in 6Li-HGY. It can be clearly seen that the average $H_2$ adsorption energy was decreased with increasing hydrogen weight percentage. For instance, at low gravimetric density (6$H_2$ in 6Li-HGY), the average adsorption energy for one $H_2$ molecule was 0.31 eV/$H_2$. With the increase of the $H_2$ molecules number per Li atom, the adsorption energy was decreased and it was reduced to 0.22 eV/$H_2$ at the maximum gravimetric density (24$H_2$ in 6Li-HGY). Not that all the average absorption energies falled in the range of -0.2 to -0.4 eV [2,3], implying the Li/HGY complex are perfect for reversible $H_2$ adsorption/desorption near room temperature.

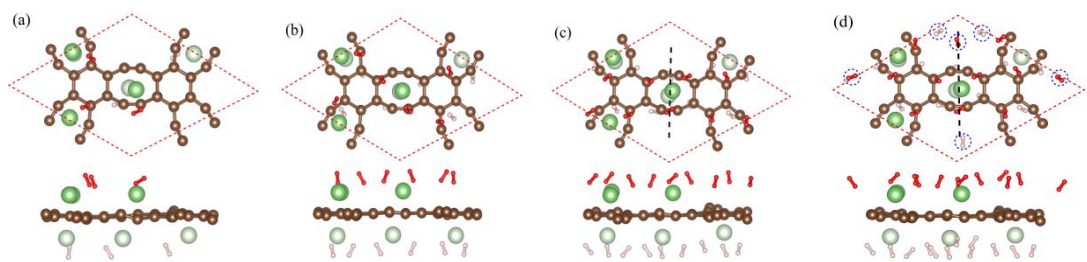

**Fig. 4.** Top and side views of atomic arrangements for (a) 6$H_2$, (b) 12$H_2$ (c) 18$H_2$ and (d) 24$H_2$ molecules adsorbed on 6Li-HGY in a unite cell. The Li and H atoms in upside and downsides of HGY sheet are plotted in different colors so as to get a better view. The black dash lines in (c) and (d) indicate one of the 2-fold rotational axes of the HGY sheet. The fourth $H_2$ molecules per Li atom at the top side of (d) are marked by blue dashed circles.

**Table 2.** Calculated Average Adsorption Energies ($E_{ads}$), Average Li-$H_2$ ($d_{Li-H2}$) Distance and H-H Bond Lengths ($d_{H-H}$) of Different $H_2$ Molecules (eV/$H_2$) on 6Li-HGY.

| System | $E_{ads}$ (eV) | $d_{Li-H2}$ (Å) | $d_{H-H}$ (Å) |
|---|---|---|---|
| 6$H_2$ in 6Li-HGY | 0.31 | 1.95 | 0.76-0.77 |
| 12H$_2$ in 6Li-HGY | 0.30 | 1.95-1.97 | 0.76-0.77 |
| 18H$_2$ in 6Li-HGY | 0.27 | 1.99-2.03 | 0.76-0.77 |
| 24H$_2$ in 6Li-HGY | 0.22 | 1.98-3.78 | 0.75-0.76 |

The PDOS of 6Li-HGY before and after H$_2$ adsorption are shown in Fig. 5a-j. A little change in the PDOS of C is found before and after the H$_2$ adsorption, which indicates that there exists a rather weak interaction between the H$_2$ molecules and the hosts materials, i.e., the acetylenic linkages of HGY sheet. While significant variation in PDOS of Li after H$_2$ adsorption can be seen, it is obvious that the H$_2$ molecules mainly interact with the Li atoms.

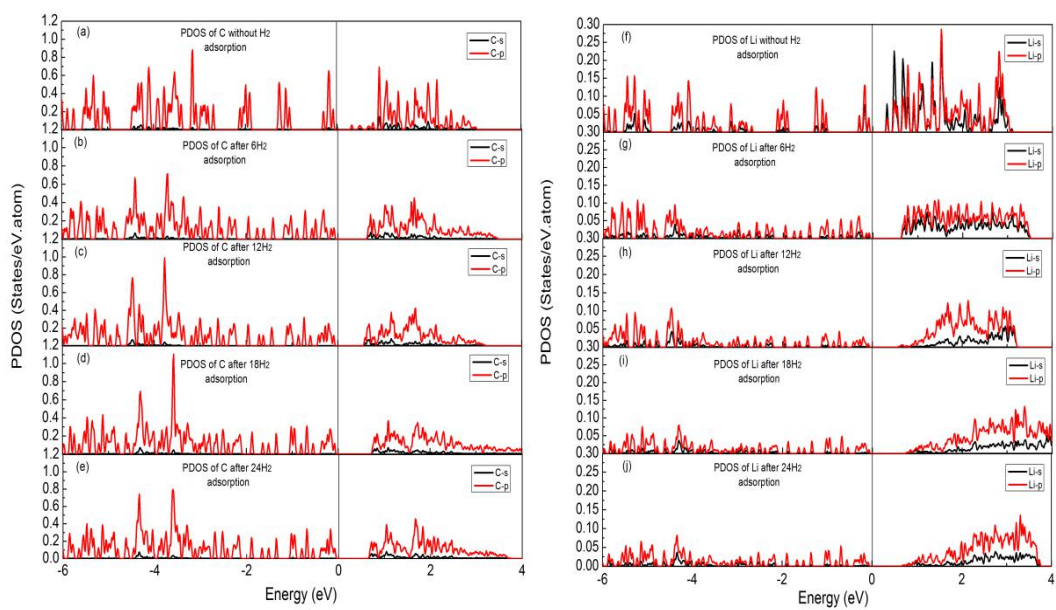

**Fig. 5.** The PDOS of C in 6Li-HGY (a) before and (b-e) after different numbers of H$_2$ adsorption, and the PDOS of Li in 6Li-HGY (f) before and (g-j) after different numbersH$_2$ adsorption.

As previously described, there are two types of hydrogen molecules at the largest hydrogen storage situation. The former connected three hydrogen molecule (1$^{st}$ type of H$_2$ molecules) are stayed close to the Li atom with the average Li-H distance of about 2.00 Å. While the forth H$_2$ molecule (2$^{nd}$ type of H$_2$ molecules) are found in relatively far from Li atoms with the average Li-H distance of about 3.77 Å, but still at the same height with other three hydrogen molecules, as illustrated in Fig. 4d.

In Fig. 6, we analyze the charge density difference of 24$H_2$ molecules adsorbed on the 6Li-HGY sheet, which would help us to reveal the interaction between the $H_2$ and the substrate sheet is whether a polarization mechanism or the weak van der Waals forces. All the $H_2$ molecules are found showing the charge depletion and accumulation, so it's clear that both the two types of hydrogen molecules bind with the substrate via the polarization mechanism. As mentioned earlier, both the HGY layer and Li ions partially charged and this produces a polarized electrostatic field, so the polarization of the 1$^{st}$ type of $H_2$ molecules is reasonable. For the 2$^{nd}$ type of $H_2$ molecules, the charge redistribution is obviously weaker than that of the 1$^{st}$ type of $H_2$ molecules as shown in Fig. 6, but what is certain is that they are also polarized. Considering the fact that the electric field of the ionic Li is almost screened by the their own polarized charges caused by the 1$^{st}$ type hydrogen molecules, we can judge that the rather weak polarization of the 2$^{nd}$ type of $H_2$ molecules is induced by the electric field provided of not the ionic Li atoms, but the polarized 1$^{st}$ type $H_2$ molecules.

To explain this polarization effect in depth, we further performed the bader charge analysis of the system. All the $H_2$ molecules were adsorbed in tilted shape so that one H atom comes close to Li atom while the second H atom is relatively far from Li atom. It's found that small amount of charge (~0.08 |e|) transferred from Li to the nearest H atoms of 1$^{st}$ type hydrogen molecule, and even smaller amount of charge (~0.05 |e|) transferred from Li to the nearest H atoms of 2$^{nd}$ type hydrogen molecule. Furthermore, the different polarization effects can also be reflected by the H-H bond lengths of the two types $H_2$ molecules, the bond lengths in the 1$^{st}$ type of $H_2$ molecules were obviously elongated and in the range of 0.758-0.764 Å, which is slightly larger than 0.750 Å of a free $H_2$. While the H-H lengths in the 2$^{nd}$ type $H_2$ molecules were only little elongated (0.752-0.753 Å), implying the rather weaker polarization compared to the 1$^{st}$ type $H_2$ molecules.

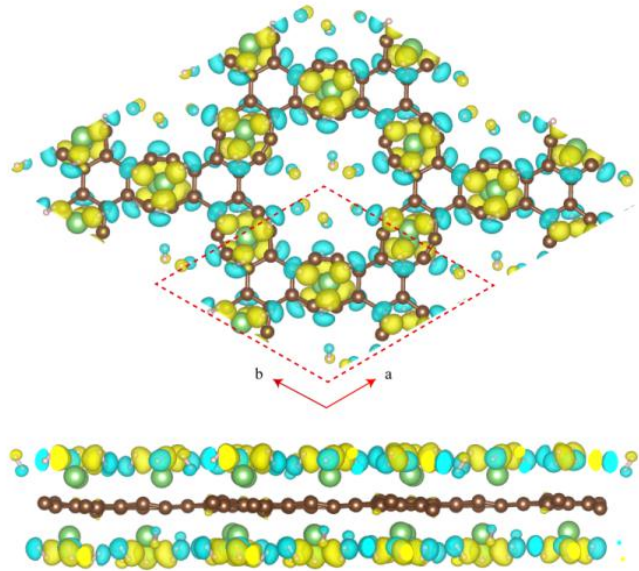

**Fig. 6.** Top and side views of the difference charge density plot induced by 24$H_2$ molecules adsorbed on 6Li-HGY Layer with the iso-surface value of $2.0\times10^{-3}$ e/Å$^3$.

## 4. Conclusions

In summary, using first-principles calculations, we investigated the hydrogen storage behavior of Li-decorated holey-graphyne. The numerical results show that such complexes can possess a hydrogen storage capacity as high as 12.8 wt%, such gravimetric density is significantly higher than the value obtained in alkali metal decorated many typical 2D materials (e.g., graphene, BN, $MoS_2$ and phosphorene). Meanwhile, the adsorption energy of per $H_2$ molecular was -0.22 eV, perfect for reversible $H_2$ adsorption/desorption near room temperature. It's revealed that the polarization mechanisms, which stems from both the electric field induced by the ionic Li decorated on the HGY and the weak polarized hydrogen molecules play a key role in the adsorption of $H_2$ molecules. Our calculations suggest that the Li atoms decorated newly synthesized holey-graphyne is a promising material as a high-capacity hydrogen storage medium.

**Acknowledgements:**

This study was supported by the Natural Science Foundation of China (Grant no.

11864033 and 11704195), the Natural Science Foundation of Shandong Province (Grant no. ZR2019MA042) and the Young Innovative Talents Fund of Shihezi University (Grant No. CXPY201906).